\documentclass[a4paper,12pt]{article}
\usepackage{cite}
\usepackage{amssymb,amsmath}
\newcommand{\be}{\begin{equation}}
\newcommand{\ee}{\end{equation}}
\newcommand{\bea}{\begin{eqnarray}}

\newcommand{\eea}{\end{eqnarray}}

\begin{document}

\title{On the ``Universal'' Quantum Area Spectrum}

\author{A.J.M. Medved \\ \\
Physics Department \\
University of Seoul \\
Seoul 130-743\\
Korea \\
E-Mail(1): allan@physics.uos.ac.kr \\
E-Mail(2): joey\_medved@yahoo.com \\ \\}

\maketitle
\begin{abstract}

There has been much debate over the form of the quantum area spectrum 
for a black hole horizon, with the evenly spaced conception of Bekenstein having featured prominently in the discourse. In this letter, we refine a 
very recently proposed method for calibrating the Bekenstein form of the spectrum. Our refined treatment predicts, as did its predecessor, a uniform
spacing between adjacent spectral levels of $8\pi$ in Planck units 
--- notably, an outcome that already has a pedigree as a proposed ``universal'' value for this intrinsically quantum-gravitational measure. Although the 
two approaches are somewhat similar in logic and quite agreeable in outcome, we argue that our version is conceptually more elegant and formally simpler than its precursor. Moreover, our rendition is able to circumvent a couple of previously unnoticed technical issues and, as an added bonus, translates to generic theories of gravity in a very direct manner.

\end{abstract}
\newpage
\section*{ }
\subsection*{Background and motivation:}

There is a long history --- beginning with the illuminating deductions 
of Bekenstein \cite{bek-1} ---
to the notion that a black hole horizon should have an inherent 
{\it quantum} area spectrum. Primarily on the basis of the 
horizon area $A$ being an 
adiabatic
invariant \cite{bek-2}, it was Bekenstein's contention that this 
spectrum
should be evenly spaced, thus leading to the famed proposal
\be
A_{n}\;=\; \epsilon l_P^2 n\;\;\;{\rm where}\;\;\;n=0,1,2,3... \; ;
\label{1}
\ee
such that $l_P^2=\hbar G_N$ is the Planck length squared, 
$n$ is the associated quantum 
number
and $\epsilon$ is some numerical coefficient of the order of 
unity. (We commit, for the time being, to four spacetime dimensions and mainly
Schwarzschild black holes, but generalizations 
will be considered in due course.) 
Let us also take note of the equivalent statement for the black
hole entropy, $S_n=\epsilon n/4$ with $n=0,1,2,...\;$, as follows
directly from the area--entropy law or $S=A/4l^2_P$ 
\cite{bek-4,haw-1}. 

Once one is willing to accept this evenly spaced form 
(which has, itself, been an issue of notable 
controversy~\footnote{Since the main criticism against the evenly 
spaced 
spectrum has 
undoubtedly come from the loop quantum gravity camp 
\cite{lqg},
it is probably worth mentioning the recent reversal in sentiment from
at least some of its supporters \cite{lqg2}.}), the remaining
point of dispute surrounds 
the spacing between adjacent levels or, equivalently, the value
of the numerical 
coefficient
$\epsilon$. There has been a persistent belief that $\epsilon$
should be equal to 4 times the natural logarithm of an integer greater
than unity \cite{bek-3}, with this being the sole means of complying 
with both
the black hole area--entropy law and a strict 
statistical
interpretation of the black hole entropy.~\footnote{To understand this
claim, one should consider both the relation $S=A/4l_P^2$ and the
usual statistical interpretation of entropy $e^{S}={\cal N}$,
with ${\cal N}$ being the number of microstates --- an integer!}
Fuel was added to this fire with
Hod's renowned conjecture \cite{hod-1} connecting $\epsilon$ to the 
spacing of the black hole
quasinormal-mode spectrum \cite{num-1,motl-1}.~\footnote{In short,
Hod proposed a direct connection between the quantum area spectrum {\it and}
the discrete but semi-classical  spectrum for the quasinormal-mode
oscillations of a perturbed black
hole. The premise for this connection is the 
{\it Bohr Correspondence Principle}, which has been conjectured 
to relate the large-$n$ limit of the former with the 
classical limit of the latter.}
At least for 
Schwarzschild black holes, this connection pinpoints the area spectral 
gap 
at precisely
$\epsilon=4\ln 3$. 

There has, nevertheless, been an alternative line of reasoning that 
suggests
a spacing coefficient of $\epsilon=8\pi$. The impetus for this brand of thought
may well have stemmed
from the current author's  observation \cite{med}  
that a number of utterly independent 
methods had all  arrived at 
the very same spacing ({\it e.g.}, \cite{kun,eg1,eg2,eg3}). 
Such diverse agreement appeared
quite unlikely to be  merely coincidental,
sparking
this author to rationalize (perhaps naively)
that $8\pi$ could indeed be a  ``universal''
value for the coefficient $\epsilon$.  But, on the 
other hand,
one truly convincing argument is better than any number
of contentious ones, and it has not been so easy to
dismiss the sound logic of Hod's exposition.
Or so it seemed until such a time when Maggiore \cite{mag}
managed to turn the whole discussion 
``on its head''.
Re-interpreting one of Hod's basic premises,~\footnote{To elucidate, Hod
regarded the classical limit of the black hole oscillation frequency
as being the real part of this frequency when the damping (or imaginary part)
goes to infinity. Alternatively, Maggiore proposed that the classical
limit is, rather, the difference in frequency moduli between adjacent
spectral levels
in the same infinite-damping limit.}
Maggiore was able to deduce
a Schwarzschild area gap that is in agreement
with the ubiquitous value $\epsilon=8\pi$. 
As a matter of fact, the very same spacing has since been confirmed
for Kerr black holes as well \cite{vag,med-2}. 

It is not our intention to dwell upon the debate between the Maggiore
and Hod interpretations. We only point out that --- if one can be swayed
to relinquish the strict statistical interpretation of black hole entropy
(as advocated in \cite{med,mag,kryp}) ---   the two
points of view would appear to be of 
comparable merit, making either of 
these a legitimate candidate for further consideration. 

The above discussion highlights the contextual backdrop for 
a very recent paper by Ropotenko \cite{rop}. That author
has furthered the intrigue with a rather novel type
of calculation for determining the Schwarzschild spectral 
spacing.~\footnote{\label{f} Although  novel, the analysis
of \cite{rop} does appear to be
conceptually similar to a 
``reduced phase-space'' program
that was initiated by Barvinsky and Kunstatter \cite{kun} 
(also see \cite{kun-2,gour-1,gour-2}). Central to both types of 
calculations is the identification of canonically conjugate
pairs in the phase space of black hole observables and
the quantization thereof. Note that the reduced phase-space method
consistently predicts a uniform spectrum with $\epsilon=8\pi$.}
Let us now recall Ropotenko's methodology, which can be summarized 
by the following paraphrased account:

One should [{\it i}] elevate the black hole ADM mass 
(or  observable energy)  
and the Schwarzschild time coordinate to quantum operators, and
then identify these
as a canonically conjugate pair,
[{\it ii}] Wick rotate from
real to imaginary time and also take the near-horizon limit
of the Schwarzschild metric
to obtain a Euclidean Rindler geometry, [{\it iii}] recognize
that 
the conjugate to the imaginary-time operator is,
up to multiplicative factors, 
one of the components (say, the $z^{\rm th}$) of an angular-momentum operator
(as follows from the well-known angular nature of
 Euclidean Rindler time \cite{sus}), 
[{\it iv}] directly relate, by way of analytic continuation, 
the {\it Lorentzian} ADM-mass operator to the newly identified
{\it Euclidean} angular-momentum operator, 
[{\it v}] utilize the first law of (black hole \cite{boys}) 
thermodynamics to translate the prior relation
into one that {\it equates} the horizon area (also elevated to 
operator status) 
to  the $z$-component of an angular momentum times a numerical
factor 
and, finally, 
[{\it vi}] apply the standard rules of quantum mechanics
to determine the area eigenvalues.

The just-reviewed analysis leads decisively to the Bekenstein spectral 
form of
eqn.~(\ref{1}) with --- as an attentive reader could well have 
anticipated --- 
a spacing of $\epsilon=8\pi$ \cite{rop}.
Moreover, the same procedure can be applied to any black hole
with a metric that can be expressed in a Schwarzschild-like form; as Ropotenko
has also explicitly demonstrated for the Reissner--Nordstrom, Kerr and 
Kerr--Newman black hole  cases. 

Although quite compelling, this calculation has a certain heuristic
feel about it; so that one might refrain 
from proclaiming it to be a rigorous 
derivation.~\footnote{Let
us be perfectly clear: The
author, Ropotenko, {\it never} did  make such a claim in \cite{rop}. The
wording has been chosen to emphasize our own personal misgivings
{\it if} such a claim is ever to be made.}
Even with such ethical concerns cast aside, there still appears to be
some unresolved 
issues of a more technical nature. These include 
[{\it a}] the viability of regarding a Wick rotation and quantization 
as commuting processes, as the author implicitly does (in the fourth step)
by claiming that the conjugate to the Schwarzschild-time operator
retains its  {\it quantum} identity as the ADM-mass operator even after
being rotated from real  to imaginary time,
and
[{\it b}]
the reliance on very specific choices of 
coordinates (Schwarzschild and its near-horizon limit)
to define the quantum operators. 
Now it is, of course, quite true that not every
physics calculation will be sensitive to the choice of coordinate system. 
However, one should probably be wary about
elevating any coordinate-dependent time parameter to the level of 
a physical observable. Indeed, this special status is normally reserved
for scalar invariants and the like.~\footnote{It should
be noted that the aforementioned  (reduced) phase-space  method 
of \cite{kun} 
faced the very same pair of issues. For this (and
any  subsequent \cite{kun-2,gour-1,gour-2})  
work, technical point [{\it a}] 
was left unresolved but clearly stated to be an assumption
of the analysis and point [{\it b}] was 
addressed 
with rigorous arguments \cite{kuch}.}

Our current motivation is, however, not really to be 
critical of the author's methodology but, rather,
to propose a strategic refinement. 
We believe that, in comparison to the prior treatment,
our version is simpler in its formalism and
more elegant in its conceptuality.
But, irrespective of anyone's own 
aesthetic preference, our rendition has
the more tangible assets 
of nicely evading the above technical caveats and of translating, 
almost directly, into
just about the most general of gravitational settings. 
Let us now proceed with an elaboration.

\subsection*{Our proposal:}

We begin here by considering the  black hole sector of
the solution space for the  Euclidean formulation of Einstein's 
theory \cite{sus}. 
When taken {\it off shell} or away from the extremum of the action, 
any such solution will typically inherit 
an {\it angular deficit}
and its associated conical singularity;
with this formal breakdown transpiring  
at the position of the ``would-be horizon''
(which is to say, at the origin
of Euclidean space in polar coordinates).
Of course, one can readily resolve this awkward state of affairs  
by simply insisting --- normally, as an  unspoken assumption ---
upon compliance with the on-shell field equations.  
Nevertheless, the off-shell solution space is still
a perfectly viable arena for asking physically meaningful questions
about just such a matter as black hole thermodynamics.

 Following along  this line of reasoning, Carlip and Teitelboim 
were, some time ago \cite{car-tei}, able to identify a physical operator 
that measures 
precisely the implicated deficit angle of 
the off-shell Euclidean theory  --- what the authors eloquently 
referred to as the 
``opening angle at the horizon'' or, in more symbolic terms, 
as $\Theta_E$.~\footnote{To be perfectly accurate, it is actually the 
quantity $2\pi-\Theta_E$ that determines the magnitude of
the angular deficit or,
equivalently, the strength of the conical singularity.  Consequently,
the angle $\Theta_E$ has an on-shell value of $2\pi$ but is
otherwise unspecified.}
More definitively, the 
so-called opening
angle can be regarded as an angular measure 
of a ``dimensionless internal time''
and can be viewed as the horizon-based analogue of the elapsed proper
time as
measured at 
radial infinity. Even more definitively,
$\Theta_{E}$ was defined  as an interval
of proper Euclidean time  elapsing
on a near-horizon surface {\it divided by}
the proper radius 
of this surface;~\footnote{In
Schwarzschild-like coordinates, for instance, the opening angle
takes on the explicit form 
$2\Theta_E=\Delta t_E\sqrt{-g_{t_Et_E}g^{rr}}/\Delta r\;$
\cite{car-tei}, with 
$\Delta t_E$ denoting the elapsed coordinate time and $\Delta r$, the  
deviation in radial coordinate distance  away from the horizon (or
away from the Euclidean origin). With this form and
the standard means of  relating
the Euclidean time periodicity to the surface gravity \cite{gib-haw},
it is straightforward
to verify that  $\Theta_E$ sweeps out an angle of exactly $2\pi$ 
during one ``orbit'' around the horizon.
Nonetheless, it should be kept in mind that the opening angle is, 
quite certainly, a coordinate-independent
construct.} so that $\Theta_{E}$ is really an angle in the most 
literal of senses.

The upshot of all this being that, after incorporating
the usually prescribed surface term for
the outer boundary \cite{gib-haw} and an  appropriately
chosen surface term  for the horizon \cite{car-tei}, one 
would find that the
off-shell version of the  Euclidean 
action $I_E$ takes on a rather telling form:
\be
I_{E}\;=\;-\Theta_E A\; +\; I_{can}\;+\;T_{E}M\; ,
\label{2}
\ee
which we have chosen to depict somewhat schematically ---
consult \cite{car-tei} for a more in-depth account.
Here, $A$ is (as before) the 
horizon
area, $T_E$ is the elapsed proper (Euclidean) time at radial 
infinity, $M$
is the ADM mass and $I_{can}$ is the usual canonical action for 
Einstein
gravity (which ultimately vanishes on shell). Also take note of the 
implied units for Newton's constant,
$G_N=1/8\pi$ or $\hbar=8\pi l_P^2$.

It should now be quite clear that, just as the ADM mass and proper time at
infinity are canonical conjugates, so too is the pair $A$ and $\Theta_E$. 
Then, since $\Theta_E$ is nothing but an angle,
it follows that its canonical conjugate --- namely, the horizon area in 
appropriate units ---
can assuredly  be identified with one of the 
components of an angular momentum. 
Alternatively, it follows directly from text-book quantum mechanics 
that the conjugate to an angle will be spectrally represented
by $\hbar$ times an integer-valued quantum 
number;~\footnote{Put simply,  
given an angle $\phi$ and its conjugate momentum
$\Pi_{\phi}$, then the eigenstate of the operator 
${\hat\phi}=-i\hbar\partial / \partial_{\Pi_{\phi}}$ --- or
$\psi_{\phi}\sim\exp[i\phi\Pi_{\phi}/\hbar]$ 
--- should be invariant
under the ``gauge'' transformation $\phi\rightarrow\phi+2\pi$.
Hence, the  spectral form of $\Pi_{\phi}$ must be an 
integral multiple of $\hbar$.} and so 
\be
A_{n}\;=\; \hbar n\;\;\;{\rm where}\;\;\;n=0,1,2,3... \; .
\label{3}
\ee
Here, as in \cite{rop}, all negative values of $n$ have been
discarded
on the grounds that the horizon interior is vanquished from Euclidean 
space.

The punch line finally comes once the units have been fully restored;
that is,
\be
A_{n}\;=\; 8\pi l_P^2 n\;\;\;{\rm where}\;\;\;n=0,1,2,3... \; 
\label{4}
\ee 
or the evenly spaced area spectrum with the advertised spacing of
$\epsilon=8\pi$.

Let us pause to reflect upon the utter simplicity 
and natural elegance  of this ``calculation''.  
Unlike the analysis of \cite{rop} --- which relied 
upon an awkward series of mathematical steps (albeit, 
each a relatively simple one) --- here, we have managed 
to {\it directly identify} the horizon area with an 
angular-momentum operator, while having used  
really no computations at all. 
Rather, some basic knowledge about the form of the 
off-shell action has sufficed for this purpose.  
It might be true that, in some sense, the math has already been done for us in
\cite{car-tei}.  Nonetheless, it is now quite 
evident that this particular manifestation of the 
horizon area is an intrinsic property as opposed to a derived one.

Let us also take this 
opportune moment to emphasize that both of the aforementioned
caveats have successfully been bypassed by our refined treatment.
First of all, the current analysis is 
restricted to the Euclidean description 
of the action --- at no time has the  analytic continuation
of a {\it quantum operator}
been invoked nor required. Secondly, one 
might have noticed that a coordinate system 
is never actually specified. This only stands 
to reason, inasmuch as our conjugate pair of observables, $\Theta_E$ and
$A$, are purely geometric entities that exist independently of
any whimsical choice of coordinates.

\subsection*{Generalizations:}

Our proposed method can be promoted further with a few pertinent observations
about  {\it generic theories of gravity}. By this terminology, we mean
$D$-dimensional  gravitational theories (whereby  $D=d+1\geq 2$)
whose respective Lagrangians are endowed with
any number  and variety of
geometric terms and/or  matter fields, 
in addition to the conventional (Einstein)  
scalar-curvature contribution.~\footnote{\label{fff}Actually, 
the gravitational Lagrangian for
such a generic theory
need not {\it explicitly} include a 
scalar-curvature
term. We do, however, insist   
upon ``sensible theories'' that 
either  reduce to or asymptotically approach
Einstein 
gravity  
in the infrared  limit.}
To be pedantically clear, a geometric term  in the Lagrangian 
is meant 
as  some  scalar-valued
functional ({\it modulo} the  $D$-``volume'' element) 
that depends on the metric, Riemann-curvature tensor and/or derivatives
thereof (as well as  possibly depending on any relevant matter field).  
Such a term could be, for instance, 
one of the ``Riemann-squared''  corrections 
of Gauss--Bonnet gravity or, perhaps, 
a representative from the so-called
$f({\cal R})$ class of theories \cite{fr}. Of course, a purely constant term
({\it to wit}, the cosmological constant) 
is also a possibility.

Let us  assume the incorporation of any boundary terms as required
to ensure a well-defined variational principle at the outer boundary
(which is no longer necessarily at radial infinity) and, if need be,
at the inner boundary (or outermost black hole horizon).   
Then, insofar as black hole solutions are admissible 
and presumed to be non-dynamical, 
the generic-gravity analogue
of the off-shell Euclidean  action for Einstein's theory
(\ref{2}) 
should be
expressible (even more schematically) as   
\be
I_E\;=\;-\Theta_E \frac{\hbar S_{W}}{2\pi}\; +\; I_{X}\;+\; B_{O}\; ;
\label{5}
\ee
where $B_{O}$ is whatever terms arise at the outer boundary,
$I_{X}$ is whatever ends up vanishing on shell
and $S_{W}$ is the well-known ``Wald (or Noether charge) entropy'' 
\cite{wald,wald-2}. 
This latter measure of {\it geometric entropy} can readily 
be interpreted as the
generic-gravity analogue of the black hole entropy 
in Einstein's theory; that is, the generalization 
of one fourth of the horizon area in Planck 
units. That
the surviving horizon term would be strictly proportional to
$S_W$ follows from the Wald entropy having been originally 
formulated 
to serve this very purpose!~\footnote{Meanwhile, that the horizon 
term is also linear
in $\Theta_E$ follows from our restriction to 
``sensible theories'' as stipulated in fn.~\ref{fff}.
This is sufficient to ensure (for a large enough black hole if absolutely  necessary) a 
leading-order Lagrangian term of the Einstein form 
${\cal R}/16\pi G_{eff}$, 
where ${\cal R}$ is the scalar curvature and $G_{eff}$ is
a ``coupling constant'' that is discussed in the very next paragraph. 
Also note that the
factor of $\hbar/2\pi$
follows from the restoration of units in
eqn.~(\ref{2}) and the area--entropy relation.} 

Another useful interpretation of the Wald entropy is supplied in
\cite{ramy}, where it has been shown that, for a
$D$-dimensional theory,
\be
S_W \;=\; \frac{A_{D-2}}{4\hbar G_{eff}}\; ;
\label{xxx}
\ee
with $G_{eff}$ denoting an {\it effective}  ``Newton's constant'' 
or a  dimensional gravitational coupling, and  
$A_{D-2}$   being regarded as the $[D-2]$-dimensional area of
a cross-section of the horizon. 
The particular relevance of this formulation being that
$G_{eff}$  --- although theory dependent --- is calculated
strictly on the  horizon and can, at least typically, be regarded
as a constant parameter for the  theory in 
question.~\footnote{The constancy of the coupling $G_{eff}$ can be viewed
as a corollary to  the so-called zeroth law of black hole 
thermodynamics, which asserts a
constant-valued surface gravity on the horizon \cite{boys}.  
Pertinently,
the surface gravity should only depend
on the gravitational coupling along with the various conserved
charges of the theory  {\it and} the zeroth law
has been  shown to hold, irrespective of the 
gravitational field 
equations, given a 
few modest impositions on the black hole solution
and global structure of the spacetime \cite{zl}.} 
From this observation, it follows that the horizon contribution
to the action (\ref{5}) is formally identical to
that of the Einstein case (a Euclidean-time parameter times a 
cross-sectional area 
divided by  a model-specific constant); meaning that the identification of
$\Theta_E$ as an angle and the rest as its conjugate will 
generically persist.   
 
Accordingly, we can, even under a remarkably general of circumstances,
anticipate the spectral form
\be
[S_{W}]_n\;=\; 2\pi n\;\;\;{\rm where}\;\;\;n=0,1,2,3... \; .
\label{6}
\ee 
Note that the same revelation was arrived at, through different means,
by Kothawala {\it et. al.} \cite{pady}.

Before concluding, let us  comment upon the added complexities for
the case of
a  charged and/or spinning black hole.~\footnote{We
will refer to a black hole's actual, physical angular momentum
as the ``spin'' to distinguish it from the angular-momentum
manifestation of the area operator.}
As
already  outlined  by Ropotenko \cite{rop},
one would generally expect the horizon-area spectral number
$n$ to be, in actuality, a composite of different quantum numbers --- 
with each such constituent number   corresponding
to one of the conserved  charges in the black hole solution; most commonly,
the mass, spin and electrostatic charge.~\footnote{If one
is interested in  complete generality,
then magnetic charge and even more exotic types of conserved  charges 
(or ``quantum hair'')  
will likely have to be included \cite{yyy}. Moreover, the spin and 
electromagnetic sectors can be expected to proliferate 
with increasing dimensionality.}
This general framework is a near certainty, given that each contributing
charge would still be subject to its individually 
prescribed process of quantization.
We would now
like to flesh out the basic premise by remarking on a few matters of
particular relevance.

Firstly, as suggested by the related ``reduced phase-space'' 
program (see fn. \ref{f}), 
one would expect any of the  constituent spectra
to be constrained by certain {\it selection rules} 
that restrict
any individual contribution (toward the composite $n$)
to a limited set of allowed values \cite{kun-2,gour-1,gour-2}.    
This happenstance is a natural consequence of the necessity for consistency 
between the area quantization rule and the standard rules
of quantum mechanics as applicable to, for instance, an Abelian charge.
Secondly, one must account for the possibility (if not the
probability)  that
two seemingly independent observables  
will no longer be decoupled  after quantization, as has been explicitly
demonstrated for the case of a   Kerr--Newman black hole in \cite{gour-2}.
There, it was shown that the electrostatic charge and spin  
are spectrally represented by a coupled system  
involving two inseparable quantum numbers. 
Finally, the reduced phase-space program strongly implies that the 
correct areal quantity to be quantized is not 
really the horizon area {\it per se} 
but, rather, 
the deviation of the area from its extremal-limiting 
value~\footnote{By which we mean
the minimal possible area for a black hole with an already
specified charge and/or spin. Reducing the mass any further, one would 
rather be describing 
a naked singularity.}   
\cite{kun-2}.
This distinction does not appear to be immediately evident in \cite{rop} or the 
current treatment but neither is it in contradiction, as one
is always principally allowed  to add (or subtract) a zero-point
value to the Bekenstein area spectrum \cite{bek-1}. To justify
this particular subtraction, one could choose to enforce it as an 
initial  condition
on the basis that 
extremal and nearly extremal black holes are highly quantum   
objects and, hence, most naturally associated with very 
small quantum numbers. 
Or, from a more classical perspective, one could also argue 
that such a subtraction  would quite sensibly
enforce {\it cosmic censorship} \cite{cc} by  
associating all naked singularities 
with  negative values of $n$.

\subsection*{Summary:}

In conclusion, we have presented yet another  method
that  both substantiates and calibrates the Bekenstein area spectrum and have
--- once again --- obtained 
the re-occurring value of $8\pi$ for the separation between adjacent levels
in Planck units.  Let us duly credit Ropotenko's recent calculation
\cite{rop} as the inspirational source for the current treatment.
We do, however, maintain that our rendition is formally simpler and
conceptually more elegant, while evading a couple of technical issues
that could well be problematic for the earlier study. 
Moreover, our analysis readily extends to generic 
theories of gravity (and matter) in an arbitrary number of
dimensions. With any luck, the ubiquitous spectral spacing of $8\pi$ 
may turn out to be ``universal" after all.


\section*{Acknowledgments}
The author's research is financially supported by the University of
Seoul.


\end{document}